# Discovery of parity-violating chiral polar-nematic charge density wave and superconductivity in kagome metals

Xingwei Shi[1,2†], Geng Li[1,2,3†,*], Zhan Wang[1,4†], Chuqi Zhang[1,2†], Ke Zhu[1,2,5], Keyu Zeng[4], Zikun Tang[5], Li Huang[1,2,3], Zhen Zhao[1,2], Jianping Sun[1,2], Xiao Liu[1,2], Jin-Guang Cheng[1,2], Chengmin Shen[1,2], Shu Ping Lau[5], Kian Ping Loh[6], Haitao Yang[1,2,3], Xiao Lin[2], Ziqiang Wang[4*] & Hong-Jun Gao[1,2,3*]

[1] *Beijing National Center for Condensed Matter Physics and Institute of Physics, Chinese Academy of Sciences, Beijing 100190, China*

[2] *School of Physical Sciences, University of Chinese Academy of Sciences, Beijing 100190, PR China*

[3] *Hefei National Laboratory, Hefei 230088, China*

[4] *Department of Physics, Boston College, Chestnut Hill, MA 02467, USA*

[5] *Department of Applied Physics, The Hong Kong Polytechnic University, Hong Kong P.R., China*

[6] *Department of Chemistry, National University of Singapore, Singapore 117543, Singapore*

**Nonmagnetic kagome metals and superconductors $AV_3Sb_5$ ($A$ = K, Rb, Cs) host unconventional charge density wave (CDW) and superconducting (SC) phases accompanied by multiple electronic symmetry breaking. Due to the centrosymmetric crystal structure, inversion symmetry has generally been assumed to hold. Here, using scanning tunneling microscopy complemented by atomic force microscopy and optical second-harmonic generation, we directly reveal that inversion symmetry in the kagome plane is spontaneously broken in the CDW state. The mixed-parity CDW state exhibits ferroelectric dipolar and nematic quadrupolar ordered moments. The coexistence and coupling between the dipole and quadrupole favor noncollinear ferro-polar and nematic alignment that breaks all mirror symmetries and gives rise to robust electronic chirality in the 3Q CDW. The multipolar coupling to in-plane electric field enables electric field control and manipulation of the chiral polar-nematic CDW state, including its chirality. Below the SC transition, we observe parity-violating pair density modulations at both the original and the CDW lattice wavevectors. Our findings of parity-violating electronic chiral multipolar order provide microscopic insights into the magnetoelectric and nonreciprocal transport, loop current order, pairing density waves, and unconventional superconductivity in kagome metals and related quantum materials.**

[*]Corresponding author. Email: hjgao@iphy.ac.cn (H.-J.G.); wangzi@bc.edu (Z.W.) gengli.iop@iphy.ac.cn (G.L.)

[†]These authors contributed equally to this work.

Due to the unique geometry of the kagome lattice, quantum interference gives rise to flat bands, Dirac crossings, and sublattice polarized single-particle electronic states (*1-5*). Many-body interactions can drive a rich landscape of novel correlated and topological quantum electronic matter (*6-11*). The recently discovered nonmagnetic kagome metals $AV_3Sb_5$ ($A$ = K, Rb, Cs) undergo a charge density wave (CDW) transition at $T_{cdw}$=78-110 K and become CDW superconductors below $T_c$~3.5 K (*12, 13*). The CDW state is highly unconventional (*14-22*) because lattice translation symmetry breaking alone is highly insufficient to describe the intriguing quantum phenomena observed, which are intertwined with correlation driven point group and other symmetry breaking. Rotation symmetry breaking appears already at elevated temperatures (*20, 23*), while a coherent electronic nematic state emerges below ~30 K (*21, 24, 25*). All mirror reflection symmetries in the kagome plane are broken, giving rise to a chiral CDW state exhibiting electronic handedness (*26-30*). Growing evidence further indicates spontaneous time-reversal symmetry (TRS) breaking (*22, 26, 31-37*) in the absence of spin-related magnetism, suggesting persistent loop-current order. In addition to the pseudogap phenomena (*15*) and Cooper pair density waves (*15, 38*), the superconducting (SC) state shows evidence for TRS breaking chiral superconductivity with residual Fermi arcs and exhibiting pair density modulations (*32*), zero-field SC diode effect (*33, 35*), and charge-6e flux quantization (*39*).

A fundamental but largely unexplored question is whether these correlated phases break inversion symmetry spontaneously even though the atomic structure is centrosymmetric. A parity-violating CDW state would be crucial for understanding the many-body origin of the CDW order, with diverse implications on the CDW chirality and manipulation, loop-current order, electric-magnetochiral transport (*28*), magnetic torque (*40*), and helical photocurrent generation (*41*). Furthermore, whether the CDW state produces parity-violating electronic states is crucial for understanding the nature of the emergent SC state and the origin of the pair density waves (*15, 38*), the nonreciprocal SC diode effect (*33, 35*), and the residual Fermi arc due to unpaired quasiparticles (*32*).

In this work, we investigate the nature of the CDW and SC states of $KV_3Sb_5$ using high-resolution scanning tunneling microscopy/spectroscopy (STM/S), supplemented by non-contact atomic force microscopy (nc-AFM) and optical second harmonic generation (SHG). By performing symmetry operations directly on real-space images. In the kagome plane, inversion symmetry is equivalent to a twofold rotation ($C_2$) and its violation would imply mix-parity electronic states. we find that the CDW state cannot be described by the commonly assumed star-of-David or inverse star-of-David configurations. While nc-AFM reveals a centrosymmetric atomic lattice, STM uncovers robust inversion-symmetry breaking driven by electronic correlation. We show that CDW state is parity-violating, comprising inversion-even and inversion-odd components, both of which break mirror symmetries and are chiral. The inversion symmetry breaking is independently corroborated by the emergence of a SHG signal below $\sim 30$ K, indicating its bulk nature. The CDW state exhibits coexisting ferroelectric polar ($\ell = 1$) and nematic quadrupolar ($\ell = 2$) order. They produce an electronic chirality and couple strongly to external electric-field, which allows manipulation of the CDW, including its chirality. Below the SC transition temperature $T_c$, spatially resolved tunneling spectroscopy reveals pronounced intra-unit-cell pair-density modulations at both $1a_0\times 1a_0$ and $2a_0\times 2a_0$ periodicities. Direct real-space inversion symmetry decomposition of the SC gap map demonstrates that the SC state also breaks inversion symmetry and contains substantial odd-parity components.

$KV_3Sb_5$ crystallizes in a centrosymmetric hexagonal structure with a stacking sequence of K-$Sb_2$-$VSb_1$-$Sb_2$-K atomic planes (Fig. 1A), and lattice parameters $a_0 = b_0 = 5.4$ Å and $c_0 = 9$ Å. The crystal cleaves between the K and the $Sb_2$ planes, exposing either a K-terminated surface or an $Sb_2$-terminated surface (fig. S1, supplementary text section 1). Similar to $CsV_3Sb_5$ and $RbV_3Sb_5$, $KV_3Sb_5$ undergoes a CDW transition around 78 K, followed by superconductivity at ~0.9 K (fig. S2). Two pronounced peaks at energies of -26 meV and 15 meV in the differential conductance spectra have been assigned to the CDW coherence peaks, indicating a CDW gap of ~ 41 meV (fig. S3). The absence of unidirectional charge order (*15, 18, 26, 29*) makes $KV_3Sb_5$ a particularly advantageous platform for studying other types of symmetry breaking electronic states.

We start by examining the normal state of $KV_3Sb_5$ at 4.2 K. nc-AFM measurements reveal the honeycomb lattice on the cleaved $Sb_2$ surface (Fig. 1B). The equal intensity of the A and B sublattices of the $Sb_2$ honeycomb lattice (line profile in Fig. 1C) indicates an intact crystal structure without distortion or buckling of the surface atoms. A zoomed-in topographic image (Fig. 1D) further confirms the perfect honeycomb lattice and the retention of inversion ($C_2$) symmetry. In sharp contrast, the STM measurements reveal distinct behaviors between the two sublattices (Fig. 1, E to G). At low bias voltages (< 40 mV), the $2a_0\times 2a_0$ CDW order, accompanied by broken rotation symmetry (*20, 26*), is clearly visible (fig. S4). The STM topography (Fig. 1E) can be well described by the triple-$\boldsymbol{Q}$ CDW with $2a_0\times 2a_0$ ($\boldsymbol{Q}_2 = \frac{1}{2}\boldsymbol{Q}_{Bragg}$) periodicity, including also the $1a_0\times 1a_0$ ($\boldsymbol{Q}_1 = \boldsymbol{Q}_{Bragg}$) charge density modulations, with rotation symmetry breaking amplitudes and phases (supplementary text section 2). Interestingly, at higher bias voltages, features associated with the CDW order are weakened (Fig. 1E, supplementary text section 2), allowing the STM topographic image to reveal two *inequivalent* sublattices (Fig. 1F), indicative of inversion symmetry breaking. A zoomed-in topographic image (Fig. 1G) clearly demonstrates the absence of inversion centers, with the $Sb_2$ atoms in the B sublattice (red triangles) consistently appearing brighter than those in the A sublattice (blue triangles).

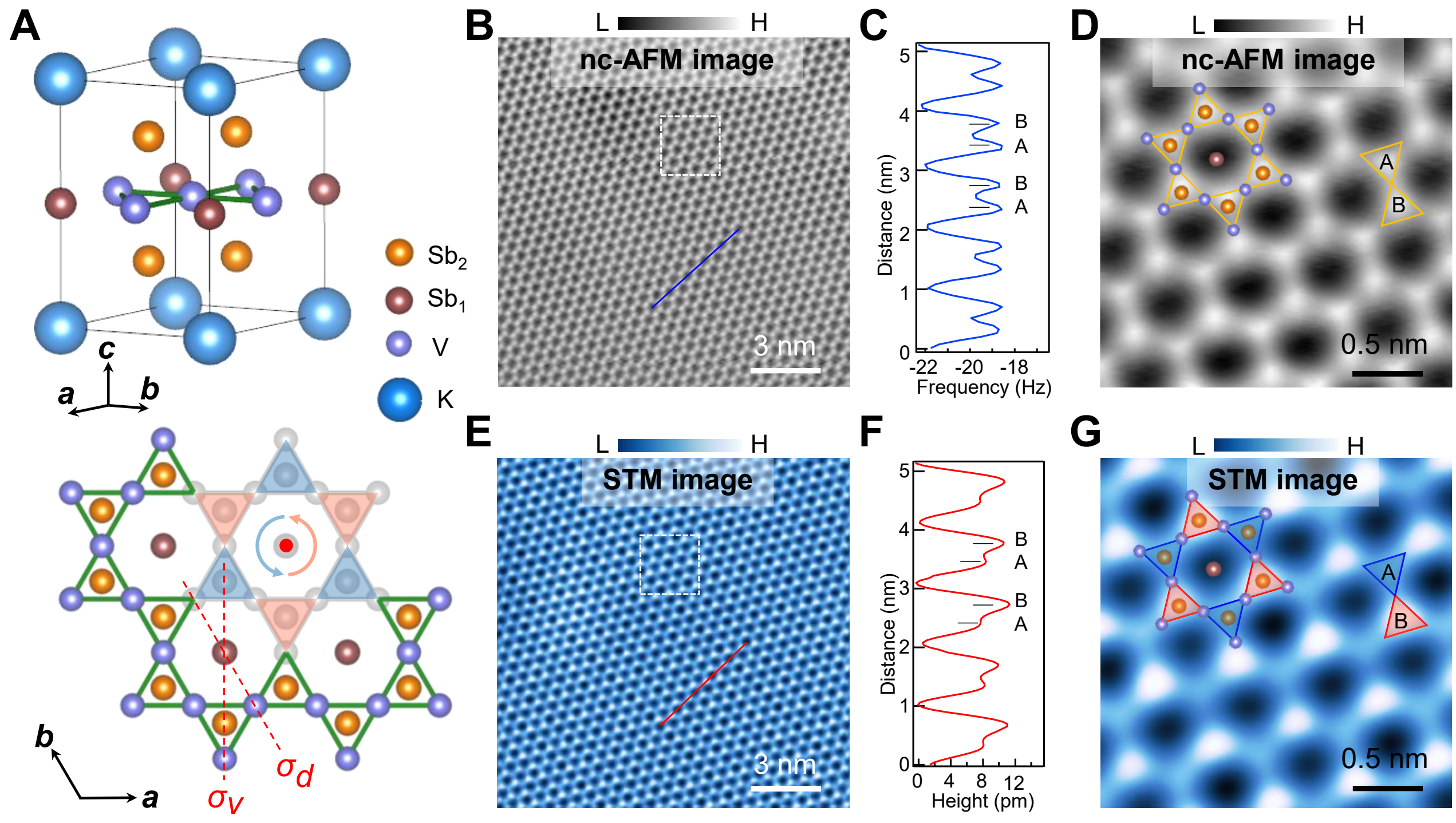

**Fig. 1. Schematic structure, AFM and STM images of the Sb surface in crystal $KV_3Sb_5$.** (**A**) 3D and top views of the crystal structure of $KV_3Sb_5$, showing the Kagome geometry of the V-Sb layer. A schematic showing the breaking of inversion symmetry of the kagome lattice is shown in the lower panel. $\sigma_v$ and $\sigma_d$ outline the two mirror axis in the kagome plane. (**B**) Large-scale constant-height nc-AFM frequency-shift image of the Sb surface of $KV_3Sb_5$. (**C**) Line profile along the blue line in (B). The frequency shifts for the two sublattices are equivalent, indicating that they are at the same height. (**D**) Zoom-in image of the region outlined by the white dashed square in (B). A star-of-David pattern with Sb and V atoms is overlaid, showing the inversion symmetry (oscillation amplitude ~100 pm, image taken at 4.2 K). (**E**) Large-scale STM topographic image of the Sb surface showing the electronic structure of the honeycomb lattice, indicating the inversion symmetry breaking of the lattice (scanning settings: $V_s$=100 mV, $I_t$=1 nA, images in (B) and (D) taken at 4.2 K). (**F**) Height profile along the red line in (E). The apparent heights of the two sublattices are different, showing the evidence of the inversion symmetry breaking. (**G**) Zoom-in image of the region outlined by the white dashed square in (E), clearly showing the inversion symmetry breaking. A star-of-David pattern with Sb and V atoms is overlaid to show the sublattice symmetry breaking and atom positions ($V_s$=100 mV, $I_t$=1 nA, images in (E) and (G) taken at 4.2 K).

We superimpose the atomic positions of the V kagome lattice containing the corner-sharing up (red) and down (blue) triangles, with the $Sb_1$ at the center of the hexagon and the $Sb_2$ above the centers of the up and down triangles (Fig. 1, D and G). The inversion operation on the kagome lattice maps the up triangle to the down triangle and IS requires them to exhibit equal brightness in the calculated STM images in the density functional theory (*42*). However, the observed STM topography (Fig. 1G) changes significantly under the inversion operation, indicating IS breaking. We find that the IS breaking topography is robust and reproducible across different samples, and the contrasting observations between nc-AFM and STM point to an electronic rather than structural origin of IS breaking.

To investigate the inversion symmetry of the CDW state, we take an STM image at a lower bias voltage in the same region as Fig. 1E, where the $2a_0\times2a_0$ CDW order is clearly resolved (Fig. 2, A and B). We first examine the topography, focusing over the region outlined by the white dashed square in Fig. 2A. A clear breaking of the $C_2$ rotational symmetry is evident in the zoom-in image (Fig. 2C) and the corresponding constant-height contour plot (Fig. 2D). By filtering out the $\boldsymbol{Q}_{CDW}$ spots in the FT image of Fig. 2B and performing an inverse FT, we isolate the real-space maps of the $2a_0\times2a_0$ CDW distribution, which reveals the prominent inequivalence between the centers of the up- and down-pointing triangles (Fig. 2E). This inequivalence is further highlighted in the constant-height contour plot in Fig. 2F. These observations indicate that the CDW state breaks inversion symmetry.

We note that inversion symmetry breaking is not visible in FT images because they artificially impose a spurious $C_2$ symmetry (*15, 20, 21*). We therefore develop a real-space symmetry decomposition (RSSD) technique (see supplementary text section 3 and fig. S5) and perform spatial inversion around the center of the "dark holes", which correspond to the "center" of the $2a_0\times2a_0$ CDW (denoted as $\boldsymbol{r_0}$ hereafter), directly on the 2D image. This allows us to disentangle even-parity and odd-parity components by adding and subtracting the two images (fig. S5). The two-dimensional (2D) inversion operation $(x, y) \rightarrow (-x, -y)$ is equivalent to the $C_2$ rotation in the kagome plane, which can in principle be studied by the real space charge-density distribution. Analyzing the inversion properties directly in real space STM images avoids a spurious $C_2$ symmetry artificially imposed in the real part of the FT images. The obtained parity-even (Fig. 2G) and parity-odd component (Fig. 2H) of the total topographic charge density modulations (Fig. 2E) show that the parity-odd part accounts for ~42 % of the total intensity of the topography, indicating significant IS breaking.

We have observed two distinct domains with opposite IS breaking patterns marked by the reversal of the relative intensities of the sublattices (fig. S6) in some regions on the same $Sb_2$ surface. Since inversion is a $Z_2$ symmetry operation, the observation of two different domains supports the broken IS. Occasionally, we also observe two different rotation symmetry breaking CDW domains (*20*) intersecting near domain boundaries. Interestingly, the relative intensities between the two sublattices remain unchanged across these boundaries (fig. S7). The observation of a single, nearly invariant inversion symmetry breaking domain across the boundaries of the CDW domains suggests that the former is independent of the rotation symmetry breaking in the CDW state.

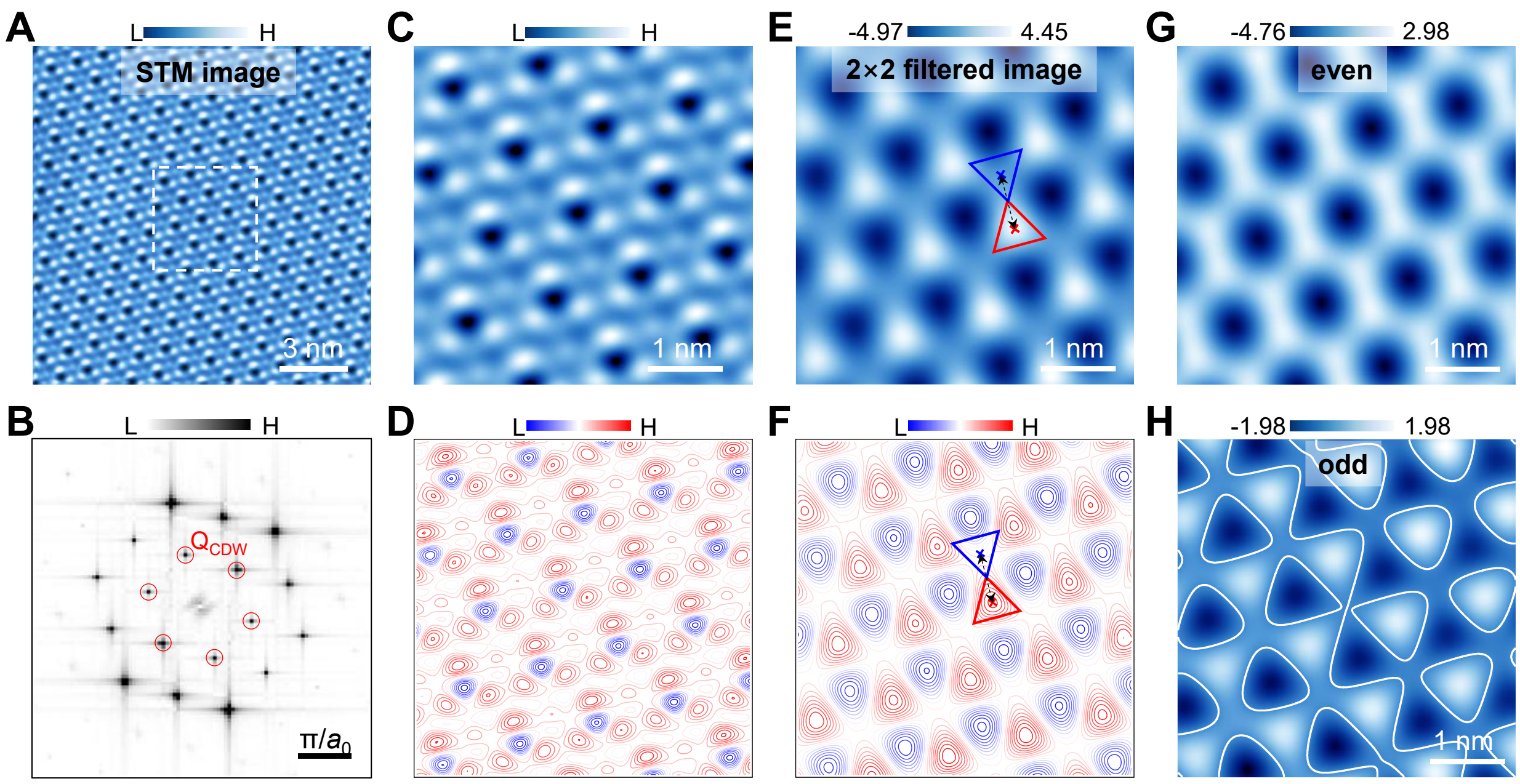

**Fig. 2. Observation of the inversion symmetry breaking in the STM images of the $KV_3Sb_5$ crystal in the CDW state at 4.2 K.** (**A** and **B**) STM image (A) and corresponding FT image (B) of the region shown in Fig. 1E**,** showing clearly the CDW state of the crystal. The $\boldsymbol{Q}_{\mathbf{CDW}}$ spots are labeled by red circles (scanning settings of (A): $V_s$=30 mV, $I_t$=1 nA). (**C**) Zoom-in image of the region in the white dashed square in (A). (**D**) Constant-height contours of (C). (**E**) Zoom-in image of the region in the white dashed square in (A)**,** after filtering the $\boldsymbol{Q}_{\mathbf{CDW}}$ spots in (B)**.** (**F**) Constant-height contours of (E). The double triangle features in (F) represent the absence of inversion centers, with the intensities in the centers of the red triangles higher than that at the centers of the blue triangles. (**G** and **H**) The even (G) and odd (H) components of (E). The white lines in (H) mark positions with zero amplitude.

To analyze the CDW order parameter in a symmetry-resolved manner and to elucidate the consequences of parity violation, we study the axipolar moment $M_\ell = \int_{u.c.} d^2r\, r^{|\ell|} e^{i\ell\theta} \rho(r,\theta)$. The integral is over the Wigner-Seitz unit cell of the topographic CDW modulation $\rho$ in 2D polar coordinates $(r,\theta)$ with the origin at the center of the unit cell, and $\ell$ labels the angular momentum (see supplementary text section 4). The statistical significance can be verified by averaging over a large number of unit cells. The axipolar analysis provides a unified and quantitative description of the parity, nematicity, multipolar and chirality of the topographic CDW modulations. Remarkably, the CDW pattern (Fig. 2A and fig. S8) is highly unconventional, as studied in detail in supplementary text section 4. The axipolar moments $M_\ell \neq 0$ in both even and odd angular moment channels, indicating robust inversion symmetry breaking. Moreover, all $M_\ell$ are complex ($M_\ell = |M_\ell|e^{i\theta_\ell}$) so that both mirror symmetries are broken (Fig. 1A), indicating handedness or chirality in both even and odd channels. In the parity-odd sector, $M_\ell$ is dominated by $\ell = 1$, producing a ferroelectric dipole moment $\boldsymbol{P} = P(\cos\theta_P, \sin\theta_P)$ with its amplitude $P = |M_1|$ and direction $\theta_P = \theta_1 = 0.29\pi$ counterclockwise away from the *x*-axis (fig. S8B). This is a ferroelectric dipole ordered (ferro-polar) CDW. Since the dipole direction is unlocked to the lattice directions, mirror symmetries are broken, resulting in the chirality $\chi_o = \mathrm{sgn}[\sin(6\theta_P)] = -1$. In the parity-even sector, $M_\ell$ is dominated by the $\ell = 2$ quadrupolar moment $Q_{ij} = Q\left(n_i n_j - \frac{1}{2}\delta_{ij}\right)$ with $Q \propto |M_2|$ and the nematic director $\boldsymbol{n} = (\cos\theta_Q, \sin\theta_Q)$ along $\theta_Q = \frac{\theta_2}{2} = -0.18\pi$ (fig. S8B), giving rise to a chiral nematic with chirality $\chi_e = \mathrm{sgn}[\sin(6\theta_Q)] = +1$. Intriguingly, the coexistence and coupling between the dipole and the quadrupole favor noncollinear ferro-polar and nematic alignment that ensures broken mirror symmetries and a robust mixed-parity electronic chirality of the 3Q CDW. The latter is described by the polar-nematic pseudoscalar $\chi_m = \mathrm{sgn}[\mathrm{Im}(M_1^2 M_2^{\bar{2}})] = \mathrm{sgn}[\sin(2\theta_P + 4\theta_Q)] = -1$, which tracks that of the chiral polar order. In supplementary text section 4.2, we show that the nematic quadrupole direction is largely pinned by the lattice, while competition between dipole–quadrupole lock-in, which enforces a near-orthogonal relation ($\theta_P - \theta_Q \approx 0.47\pi$), and dipolar lattice pinning frustrates the dipole orientation, driving it away from all mirror axes and yielding a chiral ferro-polar state.

The coexisting ferro-polar and quadrupolar nematic order implies direct coupling of the chiral CDW to external electric fields. The dipolar moment couples linearly to in-plane electric field, while the quadrupolar moment couples quadratically to electric-field analogous to electrostriction, allowing in-plane static electric-field to manipulate the multipolar and chiral properties of the CDW state (supplementary text section 4.3). Although being a dynamic probe, light-induced chirality switching of CDW has been observed recently (*29*). These findings establish inversion symmetry breaking as a central organizing principle for understanding the chiral, multipolar, and field-responsive nature of the CDW state in $KV_3Sb_5$.

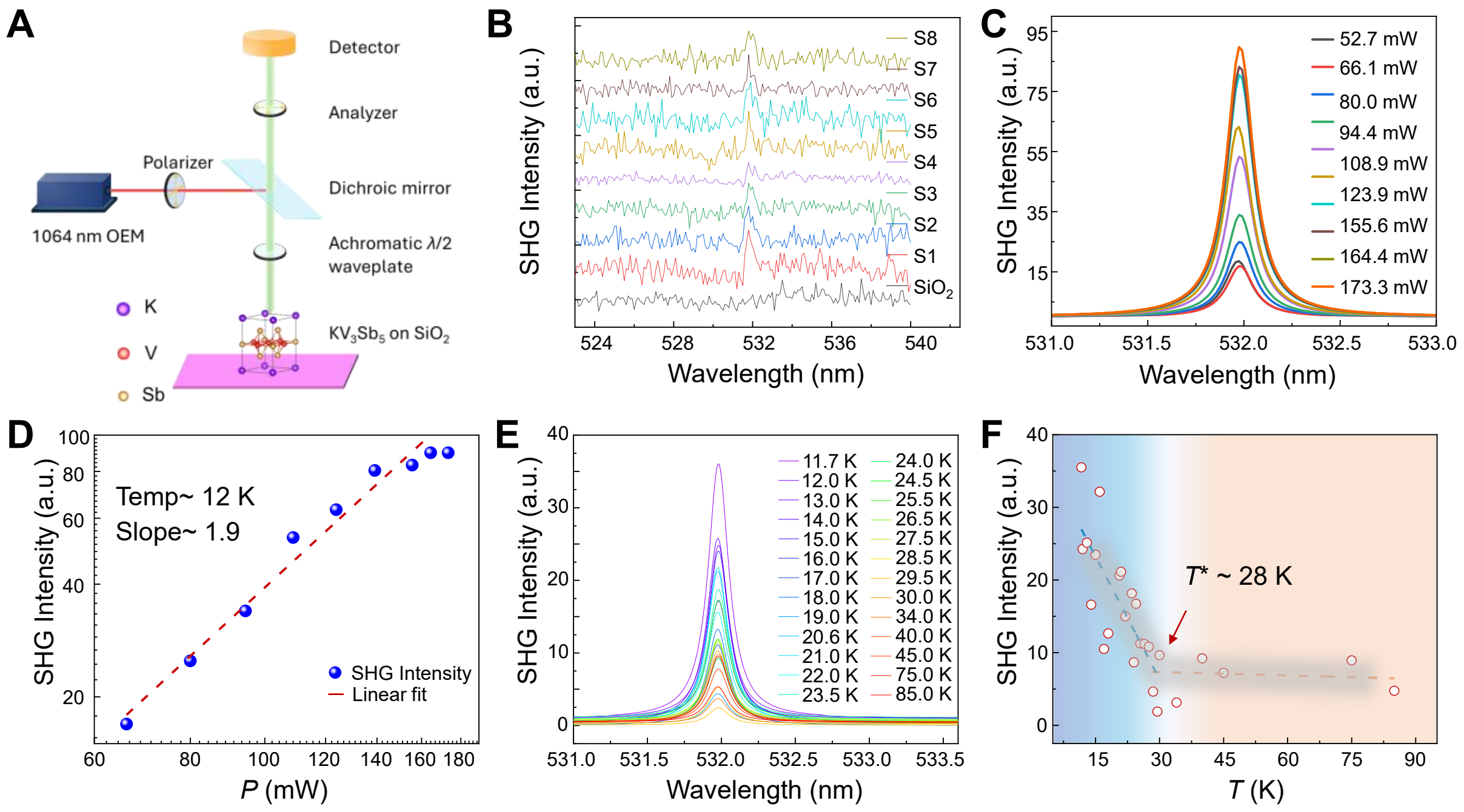

**Fig. 3. Second-harmonic generation probing of symmetry breaking in $KV_3Sb_5$.** (**A**) Schematic of the optical test setup for normal incidence SHG measurements. An excitation wavelength of 1064 nm was used. (**B**) SHG spectra of 8 different samples on $SiO_2$/Si. (**C**) Power dependence of the SHG spectra at 12 K. (**D**) SHG intensity extracted from (C). On the double-logarithmic coordinates, the linear fit yields a slope of 1.9, indicating a quadratic response. (**E**) Temperature-dependent SHG spectroscopy. (**F**) The temperature evolution of the SHG in $KV_3Sb_5$ shows a critical temperature at around 28 K.

To address whether the observed inversion-symmetry breaking is limited to the surface, we performed low-temperature second-harmonic generation (SHG) measurements. $KV_3Sb_5$ flakes were transferred onto silicon substrates for SHG experiments (Fig. 3A). At 12 K, a weak yet reproducible SHG signal is consistently observed across multiple batches of samples (Fig. 3, B-D), indicating inversion symmetry breaking in the bulk. The relatively small signal amplitude likely reflects the electronic origin of the inversion-symmetry-breaking. Notably, the SHG signal exhibits a strong temperature dependence, with a rapid increase as the temperature is lowered below~28 K (Fig. 3, E and F). The temperature evolution closely resembles that of the exotic electronic states reported to develop below ~30 K (*21, 24, 28, 36, 43*), suggesting that inversion-symmetry-breaking plays an important role in the emergent “30 K phase” of kagome metals.

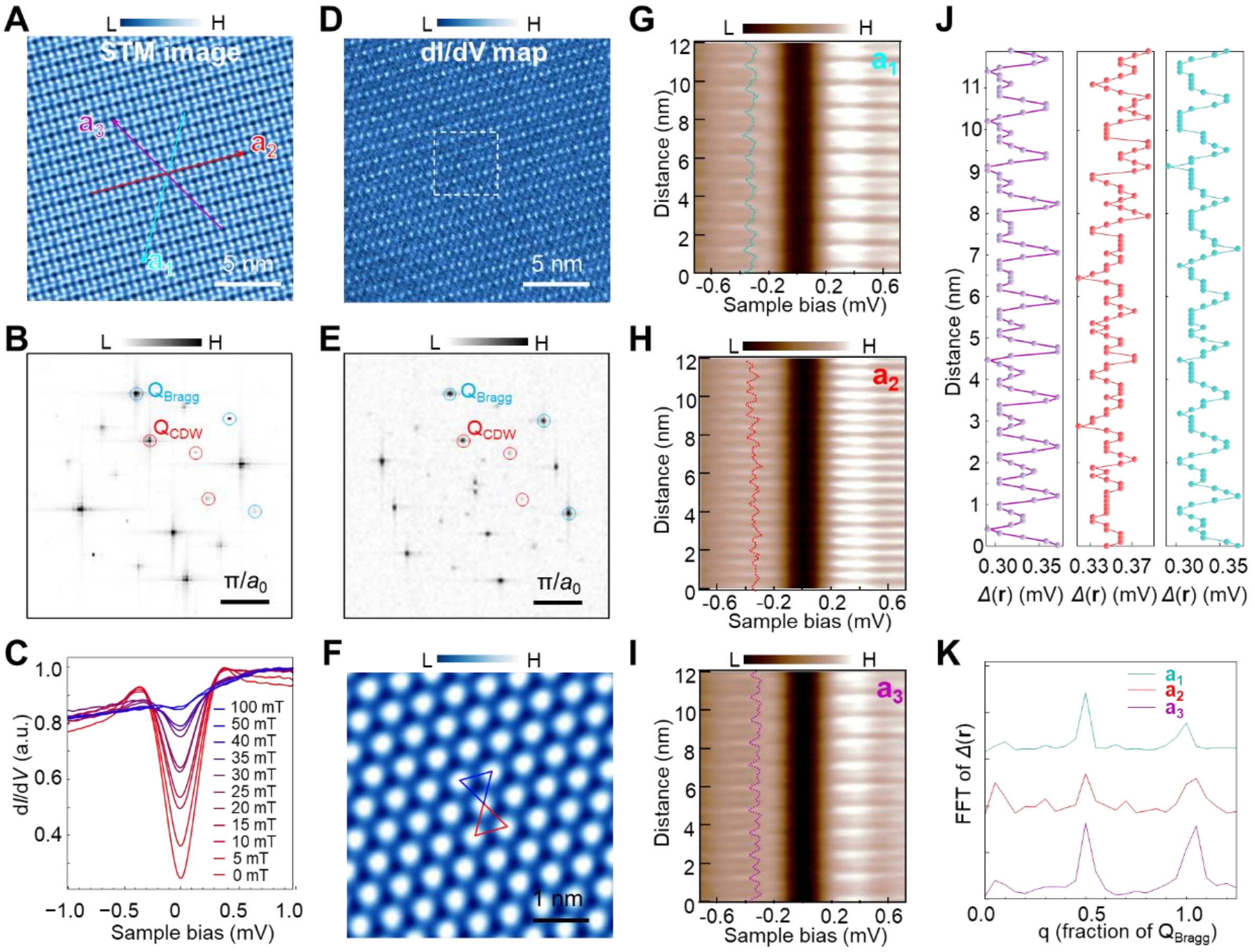

**Fig. 4. Observation of the superconducting gap modulations in the superconducting state of the $KV_3Sb_5$ crystal at 280 mK.** (**A** and **B**) Low-bias STM topographic and FT images of the Sb surface of the $KV_3Sb_5$ crystal ($V_s$=-10 mV, $I_t$=100 pA). (**C**) Field-dependent d*I*/d*V* spectra, showing the suppression of superconductivity under increasing magnetic field. The averaged superconducting gap is ~0.36 meV at 0 T. The upper critical field $Hc_2$ is determined to be ~50 mT. (**D** and **E**) d*I*/d*V* map at -0.36 mV (D) and corresponding FT image (E) of the region in (A), showing both $1a_0\times1a_0$ and $2a_0\times2a_0$ orders. (**F**) Zoom-in image of the region outlined by the white dashed square in (D), after filtering the $Q_{Bragg}$ spots in (E). (**G** to **I**) Intensity maps of the d*I*/d*V* spectra along the arrows in (A), showing superconducting gap modulations. (**J** and **K**) Superconducting gap sizes as a function of distance along the arrows (J) and corresponding FT curves (K), showing the SGM at both $Q_{Bragg}$ and $1/2Q_{Bragg}$ along different directions.

We next cool the system to ~200 mK to study the SC state. The low-bias topography and its FT image (Fig. 4A-B) show clear coexistence of CDW and superconductivity. The spatially averaged d*I*/d*V* spectra reveal a particle-hole symmetric, V-shaped SC gap of ~0.36 meV that gradually fills up with increasing magnetic field (Fig. 4C). The d*I*/d*V*(***r***, *V*) map (Fig. 4D) measured at V= -0.36 mV (left coherence peak) shows strong spatial modulations at both the CDW and Bragg wave vectors (Fig. 4E).

Remarkably, zooming-in to the outlined region in the d$I$/d$V$ map (Fig. 4D) reveals that the intensity at the center of the red triangle is higher than that at the center of the blue triangle (Fig. 4F), indicating the inversion symmetry breaking persists into the SC state.

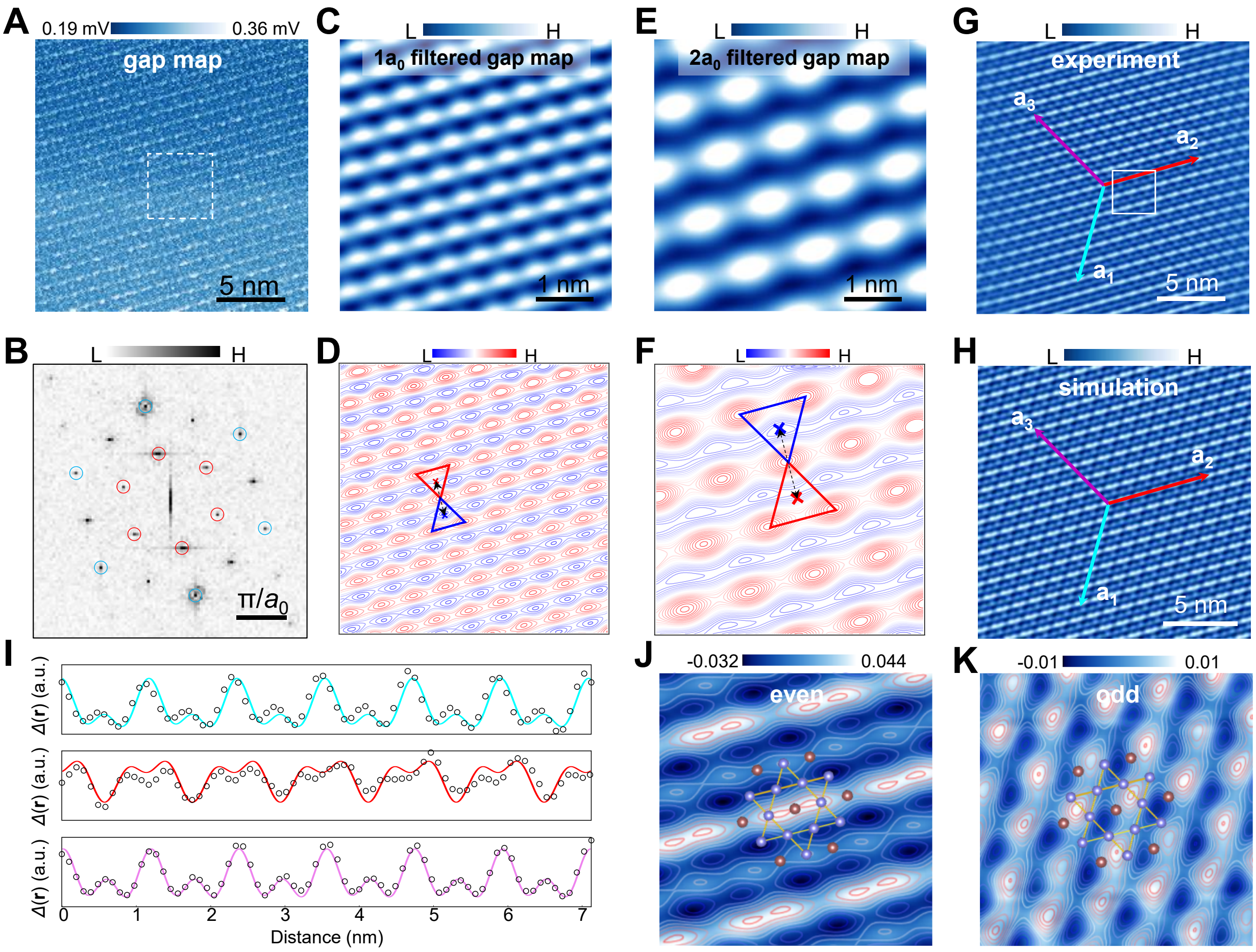

**Fig. 5. Observation of the inversion symmetry breaking in the superconducting state of the $KV_3Sb_5$ crystal at 200 mK.** (**A** and **B**) Superconducting gap map (A) and corresponding FT image (B) of the region in Fig. 4A. The $\boldsymbol{Q}_{\mathbf{Bragg}}$ and $\boldsymbol{Q}_{\mathbf{CDW}}$ spots are labeled by blue and red circles, respectively. (**C**) SC gap modulation of the white dashed square in (A), after filtering the $\boldsymbol{Q}_1$ spots. (**D**) Constant-height contours of (C), showing inversion-symmetry breaking. (**E**) The same region of (C)**,** after filtering the $\boldsymbol{Q}_2$ spots**.** (**F**) Constant-height contours of (E), showing inversion-symmetry breaking. The double triangle features represent the absence of inversion centers, with the intensities in the centers of the red triangles higher than that at the centers of the blue triangles. (**G** and **H**) Inverse FT image of the gap map in (A) by filtering the $\boldsymbol{Q}_{\mathbf{Bragg}}$ and $\boldsymbol{Q}_{\mathbf{CDW}}$ spots (G) and the simulated gap map (H), indicating IS breaking of the superconducting state. (**I**) Profile cuts along directions $\mathbf{a}_1$, $\mathbf{a}_2$ and $\mathbf{a}_3$ in (G) and (H), respectively. The open circles in **e** correspond to the experimental data from (G), whereas the solid lines are produced by the triple-Q function. (**J** and **K**) The even (J) and odd (K) components of the SGM in the region outlined by the white square in (G). The contours generated from the triple-Q function are superimposed, showing good agreement with the experimental data.

To further investigate the parity-violating SC state, we study the spatial distribution of the SC energy gap. The d$I$/d$V$($\boldsymbol{r}$, $V$) spectra along three line cuts in the high-symmetry lattice directions, labeled as $\mathbf{a}_1$, $\mathbf{a}_2$ and $\mathbf{a}_3$ in Fig. 4A, are first measured and display modulations of the coherence peaks (Fig. 4G-I). We extract the SC gap $\Delta(\boldsymbol{r})$ along the three line-cuts (Fig. 4J) and observe two sets of peaks at $1/2\boldsymbol{Q}_{\text{Bragg}}$ and $\boldsymbol{Q}_{\text{Bragg}}$ in their corresponding Fourier spectra (Fig. 4K). The SC gap modulations (SGMs) are thus associated with the wavevectors of the $2a_0$ CDW and the $1a_0$ lattice periodicity, respectively. The two sets of intra-unit-cell SGMs (*44, 45*) do not further break the lattice translation symmetry. Specifically, the $1a_0\times1a_0$ SGM is associated with $\boldsymbol{Q}_1=\boldsymbol{G}$, where $\boldsymbol{G}$ is the reciprocal kagome lattice vector. The $2a_0\times2a_0$ SGM, which has been reported in a recent study (*32*), is associated with the CDW wavevector $\boldsymbol{Q}_2=\boldsymbol{Q}_{\text{cdw}}$. The spatial modulation amplitudes of the local SC gap $\Delta(\mathbf{r})$ are about 12% of the average SC gap. The two sets of spatial SGMs are a reflection of the center-of-mass momenta of the Cooper pairs that match the reciprocal momenta of the original and CDW lattice.

To reveal the properties of the SGMs under spatial inversion, we extract the full 2D map of the SC gap (Fig. 5A) from the d$I$/d$V$ maps in the topographic region shown in Fig. 4A by identifying the SC coherence peaks at each pixel. The FT image of the gap map (Fig. 5B) clearly reveals the presence of both sets of triple-Q SGMs with $1a_0\times1a_0$ and $2a_0\times2a_0$ periodicity. To clarify the contributions from the $1a_0\times1a_0$ and $2a_0\times2a_0$ orders in the gap map, we filter the corresponding $\boldsymbol{Q}_1$ and $\boldsymbol{Q}_2$ spots in the FT images in Fig. 5B and producing the inverse FT maps back to real space. The SC gap maps are shown separately, corresponding to the $1a_0\times1a_0$ (Fig. 5C) and $2a_0\times2a_0$ (Fig. 5E) modulations, over the zoomed-in white dashed square region in Fig. 5A. Despite the prominent stripe-like features caused by rotation symmetry breaking at low bias voltages, the inequivalence between the centers of the up and down triangles is readily visible, and becomes evident in the corresponding constant-height contour plots in Fig. 5D and Fig. 5F. We note that the observation is reproducible in different samples (fig. S9).

These results provide direct evidence for the parity-violating superconductivity by revealing that the spatial distribution of the SC pairing energy gap breaks inversion symmetry. Both $1a_0\times1a_0$ and $2a_0\times2a_0$ SGMs in the gap map (Fig. 5G) can be well described quantitatively by a triple-Q SC gap function (Fig. 5H, supplementary text section 5). We plot in Fig. 5I the SGMs along the three high-symmetry directions represented by the $\mathbf{a}_1$, $\mathbf{a}_2$ and $\mathbf{a}_3$ vectors marked in Fig. 5G and 5H, showing the remarkable agreement with the triple-Q SGM function with both the $1a_0$ and $2a_0$ periodicities. Importantly, we find that the relative phases of the triple-Q SGMs are very different from those carried by the triple-Q

CDW (supplementary text section 4), which strongly supports that the SGMs are intrinsic to SC pairing in kagome superconductors.

We further apply the same RSSD analysis developed for the CDW state to the SGM map in Fig. 5G, enabling a decomposition into even- and odd-parity components (Fig. 5 J-K). The registry of the atomic positions on the spatial distribution of the SGMs is discussed in supplemental text section 5, with all in-plane $Sb_1$ atoms residing on the nodal lines of the odd-parity component (Fig. 5K, fig. S10 and fig. S11). The comparison between the intensity in Fig. 5J and 5K reveals that a substantial fraction (~23%) of the SGMs arise from the odd-parity component (Fig. 5, J-K and fig. S10), providing direct and robust evidence of inversion-symmetry breaking in the SC state. We note that this estimate represents a lower bound for odd-parity pairing, since a uniform odd-parity component of the condensate would not contribute to the spatial modulation. The close correspondence between the RSSD analysis of the CDW and SGMs reveals a chiral polar-nematic SC ground state for the kagome superconductors.

It is instructive to note that inversion symmetric $2a_0 \times 2a_0$ pair density modulation (*46*) and coexisting $1a_0 \times 1a_0$ and $2a_0 \times 2a_0$ pair density modulations (*47*) have been shown recently to appear in theoretical studies of the SC state emerging from an inversion symmetry CDW state with loop-current order. These are inconsistent with our observed parity violating SC pairing density modulations. This discrepancy highlights the fundamental importance of parity violation in the CDW state, which has not been studied to date. Moreover, since the pairing takes place in the inversion-symmetry breaking electronic structure of the CDW normal state, the SC state in general must involve mixed-parity pairing containing both spin-singlet and spin-triplet pairing components.

In summary, we reported the discovery of parity-violating CDW and SC states in kagome metals, despite the centrosymmetric crystal lattice. Previous studies have focused a parity-even bond-ordered CDW, typically described by the Star-of-David or inverse Star-of-David configurations (*42*). Different stackings of the CDW configurations along the c-axis, which can induce rotation symmetry breaking (*42, 48-51*), cannot lead to inversion symmetry breaking. Thus, our observation of the absence of spatial inversion centers in the electronic states reveals that inversion symmetry in the kagome plane is *spontaneously* broken due to electronic correlations. We performed mean-field theory calculations using the minimal one-orbital model on the kagome lattice and found that strong intersite Coulomb interaction can indeed produce an inversion-symmetry breaking mixed parity CDW state with $2a_0 \times 2a_0$ bond and charge distributions consistent with the topography observed by STM (supplementary text section 5, fig. S12). The observation of the novel ferroelectric polar-nematic CDW and superconductivity not only provides the microscopic origin for the chiral electronic matter, but also opens new paths for electric field and lattice strain control and manipulation of the electronic chirality. They have immediate implications for the time-reversal symmetry breaking loop-current order, electric-magnetochiral and nonreciprocal transport, and the unconventional chiral superconductivity. Our findings of the parity-violating quantum electronic states as a central organizing principle are transformative for exploring the correlated and topological matter in kagome metals and related quantum materials.

## References

1. E. Tang, J. W. Mei, X. G. Wen, *Phys. Rev. Lett.* **106**, 236802 (2011).
2. K. Sun, Z. C. Gu, H. Katsura, S. Das Sarma, *Phys. Rev. Lett.* **106**, 236803 (2011).
3. M. L. Kiesel, R. Thomale, *Phys. Rev. B* **86**, 121105 (2012).
4. W. S. Wang, Z. Z. Li, Y. Y. Xiang, Q. H. Wang, *Phys. Rev. B* **87**, 115135 (2013).
5. H. M. Guo, M. Franz, *Phys. Rev. B* **80**, 113102 (2009).
6. L. D. Ye *et al.*, *Nature* **555**, 638-642 (2018).
7. J. X. Yin *et al.*, *Nature* **562**, 91-95 (2018).
8. E. K. Liu *et al.*, *Nat. Phys.* **14**, 1125-1131 (2018).
9. N. Morali *et al.*, *Science* **365**, 1286-1291 (2019).
10. M. G. Kang *et al.*, *Nat. Mater.* **19**, 163-169 (2020).
11. J. X. Yin *et al.*, *Nature* **583**, 533-536 (2020).
12. B. R. Ortiz *et al.*, *Phys. Rev. Materials* **3**, 094407 (2019).
13. B. R. Ortiz *et al.*, *Phys. Rev. Lett.* **125**, 247002 (2020).
14. S. Y. Yang *et al.*, *Sci. Adv.* **6**, eabb6003 (2020).
15. H. Chen *et al.*, *Nature* **599**, 222-228 (2021).
16. S. Zhou, Z. Q. Wang, *Nat. Commun.* **13**, 7288 (2022).
17. H. S. Xu *et al.*, *Phys. Rev. Lett.* **127**, 187004 (2021).
18. H. Zhao *et al.*, *Nature* **599**, 216-221 (2021).
19. M. G. Kang *et al.*, *Nat. Phys.* **18**, 301-308 (2022).
20. H. Li *et al.*, *Nat. Phys.* **18**, 265-270 (2022).
21. L. P. Nie *et al.*, *Nature* **604**, 59-64 (2022).
22. C. Mielke *et al.*, *Nature* **602**, 245-250 (2022).
23. Y. Xiang *et al.*, *Nat. Commun.* **12**, 6727 (2021).
24. H. Li *et al.*, *Nat. Phys.* **19**, 637-643 (2023).
25. C. Guo *et al.*, *Nature* **647**, 68–73 (2025).
26. Y. X. Jiang *et al.*, *Nat. Mater.* **20**, 1353-1357 (2021).
27. N. Shumiya *et al.*, *Phys. Rev. B* **104**, 035131 (2021).
28. C. Y. Guo *et al.*, *Nature* **611**, 461-466 (2022).

29. Y. Q. Xing *et al.*, *Nature* **631**, 60-66 (2024).

30. H. J. Elmers *et al.*, *Phys. Rev. Lett.* **134**, 096401 (2025).

31. T. Neupert, M. M. Denner, J. X. Yin, R. Thomale, M. Z. Hasan, *Nat. Phys.* **18**, 137-143 (2022).

32. H. B. Deng *et al.*, *Nature* **632**, 775-781 (2024).

33. T. Le *et al.*, *Nature* **630**, 64-69 (2024).

34. Y. Xu *et al.*, *Nat. Phys.* **18**, 1470-1475 (2022).

35. J. Ge *et al.*, *arXiv:2506.04601*, (2025).

36. H. Gui *et al.*, *Nat. Commun.* **16**, 4275 (2025).

37. J. Huang *et al.*, *arXiv:2512.11341*, (2025).

38. X. H. Han *et al.*, *Nat. Nanotechnol.* **20**, 1017-1025 (2025).

39. J. Ge *et al.*, *Phys. Rev. X* **14**, 021025 (2024).

40. T. Asaba *et al.*, *Nat. Phys.* **20**, 40-46 (2024).

41. Z. J. Cheng *et al.*, *Nat. Commun.* **16**, 3782 (2025).

42. H. X. Tan, Y. Z. Liu, Z. Q. Wang, B. H. Yan, *Phys. Rev. Lett.* **127**, 046401 (2021).

43. C. Guo *et al.*, *Nature* **647**, 68-73 (2025).

44. L. Y. Kong *et al.*, *Nature* **640**, 55-61 (2025).

45. T. H. Wei *et al.*, *Chinese. Phys. Lett.* **42**, 027404 (2025).

46. M. Yao, Y. Wang, D. Wang, J. X. Yin, Q. H. Wang, *Phys. Rev. B* **111**, 094505 (2025).

47. Z. Wang, K. Y. Zeng, Z. Wang, *arXiv:2504.02751*, (2025).

48. M. H. Christensen, T. Birol, B. M. Andersen, R. M. Fernandes, *Phys. Rev. B* **104**, 214513 (2021).

49. R. Gupta *et al.*, *Communications Physics* **5**, 232 (2022).

50. Y. Hu *et al.*, *Phys. Rev. B* **106**, L241106 (2022).

51. M. G. Kang *et al.*, *Nat. Mater.* **22**, 186-193 (2023).